\documentclass[%
 reprint,
 amsmath,amssymb,
 aps,
]{revtex4-2}

\usepackage{algorithm}
\usepackage{algpseudocode}
\usepackage{graphicx}
\usepackage{dcolumn}
\usepackage{bm}
\usepackage{comment}

\usepackage{subcaption}
\usepackage{hyperref}

\newcommand{\indx}{\nu}
\newcommand{\Tunit}{\mathcal{E}/k_B}

\begin{document}

\preprint{APS/123-QED}

\title{Efficient kinetic Monte Carlo simulations with long-range electrostatic interactions}

\author{Roya Ebrahimi Viand}
\email{viand@fhi-berlin.mpg.de}
\affiliation{Fritz-Haber-Institute of the Max Planck Society, Faradayweg 4-6, 14195 Berlin, Germany}

\author{Bat-Amgalan Bat-Erdene}
\affiliation{Fritz-Haber-Institute of the Max Planck Society, Faradayweg 4-6, 14195 Berlin, Germany}

\author{Karsten Reuter}
\affiliation{Fritz-Haber-Institute of the Max Planck Society, Faradayweg 4-6, 14195 Berlin, Germany}

\author{Sebastian Matera}
\affiliation{Fritz-Haber-Institute of the Max Planck Society, Faradayweg 4-6, 14195 Berlin, Germany}
\affiliation{Institute for Mathematics, Freie Universit\"at Berlin, Arnimallee 6, D-14195 Berlin, Germany}



\begin{abstract}
Charge transport in solid-state materials is often governed by charge carrier hopping processes in the presence of long-range electrostatic interactions. Kinetic Monte Carlo (kMC) simulations provide a framework for describing such rare-event dynamics over extended timescales. However, the efficient treatment of long-range interactions remains a major computational challenge, since each particle jump modifies the energy landscape globally and, in principle, requires updating all transition rates after every carrier motion.
We present an efficient and generally applicable update process for these transition rates in the presence of long-range electrostatic interactions. 
To illustrate the method, we study charge diffusion on a simple cubic host lattice under an external electric field.
The transport behavior is investigated in high states of charge (SOC), and the influence of temperature, electric field strength, and SOC is examined. Our simulations show strongly suppressed transport at 100\,\% SOC (50\,\% occupation) caused by a freezing of the charge carriers into a Coulomb superlattice. Slight deviations from this reference in terms of charge carrier concentration lead to  a rapid increase in conductivity. A deeper analysis reveals that this behavior can
be rationalized as transport of non-interacting defects in the Coulomb superlattice.
These results demonstrate the capability of the proposed update procedure to efficiently capture non-equilibrium transport phenomena in interacting charged systems and provide a foundation for simulations of more complex charge diffusion problems.
\end{abstract}

\maketitle


\section{Introduction}

Charge transport in solid materials is often controlled by the interplay of rare diffusive jumps of (quasi-)classical charge carriers, e.g. the diffusion of ions in solid electrolytes \cite{Bachman2016} or localized polarons in semiconductors \cite{yu2005fundamentals, Franchini2021}. The rare event nature prohibits the use of methods like molecular dynamics (MD) which fully resolve the atomic motion and suffer from a large timescale mismatch between fast lattice vibrations and infrequent jumps, resulting in impractically long simulation times for observing diffusion events. Kinetic Monte Carlo (kMC) simulations (also known as Dynamic Monte Carlo, Stochastic Simulation Algorithm, Event Driven Monte Carlo or Gillespie Method) address these challenges by coarse-graining the dynamics to a rare jump description and simulating the resulting Markov process \cite{Voter2007, Voter2002, vanderVen2013, Ruhle, 10.1021/ar900119f}, thereby providing access to timescales beyond what can be reached by atomistic simulations \cite{Jassar2025, kmc_canepa, Harald_oberhofer_2017}.

A fundamental challenge in the field is to understand charge carrier transport and its computational treatment with kMC in the high state of charge (SOC) where long-range Coulomb interactions play an essential role\cite{Maier2005}. Typical kMC simulations involve local updates and, in each step, only $\mathcal{O}(1)$ particles are moved. Without long-range interactions, this can be exploited to achieve a complexity per step of $\mathcal{O}(\log N)$ (when $N$ is the system size) or even $\mathcal{O}(1)$ complexity when utilizing symmetries\cite{HOFFMANN20142138}. In contrast,  long-range interactions require all process rates to be updated and the complexity changes to $\mathcal{O}(N \log N)$ per step, e.g., using Particle-Mesh-Ewald \cite{particle_mesh}. As a consequence, the relative impact of long-range interactions is significantly more pronounced in kMC than in methods like MD, where the complexity only changes from $\mathcal{O}(N)$ to $\mathcal{O}(N \log N)$. Efficient update schemes are thus essential for the kMC simulation of charge transport.

To date, existing kMC implementations often rely on simplifications to manage the computational cost of long-range Coulomb interactions, such as interaction cutoffs, limiting interactions to near-neighbor environments, or precomputed rates for limited configurations\cite{kmc_cutoff_weakness, VANDERKAAP2016321, kmc_charge_package_2026, DENG_kmc_package}.
Alternative approaches include cluster expansion–based frameworks that describe configurational energetics using local interactions \cite{vanderVen2001}, as well as schemes that separate short- and long-range contributions and treat the latter at a mean-field level \cite{SZABO_coloumb_gas}. For cutoff-based treatments, various strategies have been proposed to improve the accuracy of the truncated Coulomb interaction \cite{efficient_coloumb_kmc_2020}; however, truncating or approximating long-range interactions will, in principle, influence the resulting transport behavior \cite{kmc_cutoff_weakness}.
Without these simplifications, the efficient update of process rates after each particle jump remains the principal challenge, scaling with system size due to the long-range nature of the interactions. In this work, we introduce a generally applicable and computationally efficient update procedure for kMC simulations with long-range interactions. The proposed procedure enables the exact treatment of full-range Coulomb interactions and can be applied to arbitrary lattice structures and multiple interacting species. This is achieved by exploiting the mathematical structure of typical rate expressions mapping the long-range update onto a vector-vector multiplication. While this still comes at a complexity of $\mathcal{O}(N)$, it can be implemented very efficiently on modern compute architectures.

To illustrate the approach, we consider a specific example for diffusion of a single species on a simple cubic host lattice under an external electric field. In particular, we investigate the behavior close to $100$\,\% state of charge and test for the influence of charge carrier concentration, temperature, and electric field strength on the non-equilibrium transport. This is of particular interest in the context of battery materials, where charge densities close to $100$\,\% SOC commonly appear, e.g., at the end of a charging cycle. We find that the current drops significantly as we approach $100$\,\% SOC which is caused by the freezing of charge carriers into a Coulomb superlattice. We rationalize the transport in this regime in terms of non-interacting defects in this superlattice. Defining the defect concentration to be proportional to the deviation of the charge carrier concentration from the $100$\,\% SOC, an analytical ultra-dilute defect model leads to a perfect fit with our simulation data.

\section{Theory}

The diffusion of quasi-classical charge carriers (like ions or localized polarons) in solids is characterized by long periods of atomic vibrations around equilibrium positions and rare jumps to neighboring sites. These jumps require overcoming substantial energy barriers, often in the range of 0.15-3 eV \cite{Wang2015, Marino_NIAL}. The large separation between vibrational timescales (femtoseconds) and diffusive jumps (nanoseconds to milliseconds) makes the transport a rare event problem and an ideal system for kMC simulation \cite{Ruhle, 10.1021/ar900119f}.
Due to this timescale separation, we can treat diffusion as a Markov process where the probability of a particle jump depends only on the current configuration, not on system history. This is justified by the thermalization that occurs between rare jumping events, effectively erasing memory of previous states \cite{Voter2002}.
We consider charge carriers moving on a fixed lattice with $N$ lattice sites provided by a rigid host structure. The host lattice defines the available sites and possible hopping processes, while its atomic degrees of freedom are not explicitly propagated. This rigid-host approximation is appropriate when structural changes of the host are small on the timescale of  carrier diffusion or when local host relaxations can be incorporated into effective hopping barriers and transition rates.
The system state in a general lattice model is represented by an integer vector $x$ encoding which species occupies which individual site. With obvious generalizations, we assume that each site can only be occupied by a single particle. This allows us to employ a binary representation. In this representation, each entry $x_i$ corresponds to a certain site and a single particle species. The site is occupied by a particle of this species when $x_i=1$ and else $x_i=0$. Note, in the case of multiple species, multiple entries can correspond to the same site but will encode the occupation of this site by different species. 

With the state representation established, we can express the time evolution of the probability distribution $P(x,t)$ through the master equation (ME)
\begin{equation}\label{eq:master_equation_traditional}
\frac{dP(x,t)}{dt} = \sum_{x'} \left[ W(x|x')P(x',t) - W(x'|x)P(x,t) \right],
\end{equation}
where $W(x|x')$ represents the transition rate from state $x'$ to state $x$. The first term describes probability flowing into the state $x$ from all other states, while the second term accounts for probability flowing out of the state $x$ to other states.

\begin{figure}
\includegraphics[width=\columnwidth]{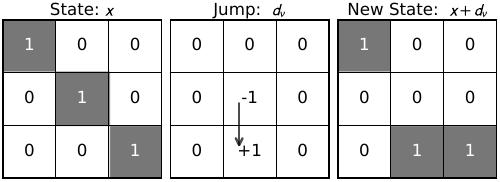}
\caption{\label{fig:transition_vector}
The left panel shows the original state of the system $x$, where gray cells indicate occupied sites and white cells represent empty sites. The middle panel illustrates the process vector $d_{\indx}$, being made up of $0$ values except for the destination and the departure sites. The right panel displays the new state $x + d_{\indx}$ after the process occurs. }
\end{figure}

For discussing algorithms in the following, it is more convenient to reformulate Eq. \eqref{eq:master_equation_traditional} in terms of processes. A process $\indx$ can be regarded as the collection of transitions $x' \rightarrow x$ for which $x-x'= d_\indx$, where $d_\indx$ is a fixed  (sparse) integer vector representing the change in occupation. This change is typically local, i.e. only the entries corresponding to sites within a certain local environment are changed by the process $\indx$ and therefore $d_\indx$ has only $\mathcal{O}(1)$ non-zero entries. 
Figure~\ref{fig:transition_vector} provides a 2D illustration of this state vector transformation for a special case of a square lattice with only one particle type and diffusive jumps of single particles between neighboring sites. It begins with the initial configuration $x$, where occupied sites are shown in gray. The process vector $d_\indx$ encodes the departure (-1) and arrival (+1) sites of the moving particle and results in the final configuration $x + d_\indx$ after the process execution.
Each process has its associated propensity function $a_{\indx}(x):=W(x+d_{\indx}|x)$ (also known as rate function, reaction rate, or intensity function). This leads to the process-oriented form of the master equation
\begin{equation}\label{eq:master_equation}
\frac{dP(x,t)}{dt} = \sum_{\indx=1}^{N_\text{proc.}}\left[ a_{\indx}(x-d_{\indx})\,P(x-d_{\indx},t) - a_{\indx}(x)\,P(x,t) \right],
\end{equation}
where $N_\text{proc.}$ is the total number of processes. $N_\text{proc.}=\mathcal{O}(N)$ holds when processes are local.

\subsection{Kinetic Monte Carlo}
Direct solution of the master equation is generally intractable for large systems due to the curse of dimensionality. The number of coupled differential equations grows exponentially with system size. As an example, a $10\times 10\times 10$ lattice that contains one particle type, requires solving $2^{1000} > 10^{300}$ coupled ODEs, which exceeds  any current or foreseeable computational capabilities. While specialized mathematical approaches such as tensor train approximations can provide solutions for master equations with particular mathematical structure \cite{Gelss_Matera_2015}, most realistic systems require sampling-based approaches exploiting that the ME describes a stochastic process $X(t)$. Instead of simulating the ME \eqref{eq:master_equation}, we simulate sample trajectories of the process $X(t)$, which is formally given by the random time change representation \cite{Kurtz1981}
\begin{equation}
X(t) = X(0) + \sum_{\indx=1}^{N_{\text{proc.}}} d_{\indx}\, Y_{\indx}\left( \int_{0}^{t} a_{\indx}(X(s))\,ds \right),
\label{eq:kurtz}
\end{equation}
where $Y_{\indx}(*)$ are unit rate Poisson processes. This pathwise representation forms the theoretical basis for the kMC method, which generates sample trajectories of the Markov process defined by Eq. \eqref{eq:kurtz}. Observables are then estimated as averages over these stochastic realizations, which reproduce the statistical properties governed by the master equation given in \eqref{eq:master_equation} \cite{Gillespie1976,Bortz1975}.

\begin{algorithm}[H]
\caption{Direct Method Algorithm \label{direct_algorithm}}
\begin{algorithmic}[1]
\Repeat
    \State Update propensity values $a_\indx(x)$ for the current state $x$
    \State Calculate total propensity $a = \sum_\indx a_\indx(x)$
    \State Generate random number $r_1 , r_2 \in (0,1]$
    \State Calculate time increment $\Delta t = -\ln(r_1)/a$
    \State Update time $t \leftarrow t + \Delta t$
    \State Select process $\indx$ such that $\sum_{\nu=1}^{\indx -1} a_\nu(x) < r_2 a \leq \sum_{\nu =\indx}^{N_\text{proc.}} a_\nu(x)$
    \State Update the state: $x \rightarrow x + d_\indx$
\Until{simulation complete}
\end{algorithmic}
\end{algorithm}

There are different kMC algorithms available, mostly based on two primary classes of rejection-free kMC algorithms, both originally developed by Gillespie\cite{Gillespie1976}. The difference between the two methods is how they determine when and which process to execute next. The first class, which we use, is the Direct Method shown in algorithm \ref{direct_algorithm}. This algorithm uses two random numbers, one to determine the time step and another to choose the next process based on their relative likelihoods. The other method, named as First Reaction Method, assigns a random firing time to each possible event and executes the one that occurs soonest. Both approaches are equivalent and there is no natural preference for one of them. Rather, the performance of an actual kMC code depends more on the employed data structures and the actual implementation.

\subsection{Propensities and Fast Update Rules}
\begin{figure}
    \includegraphics[width=0.7\columnwidth]{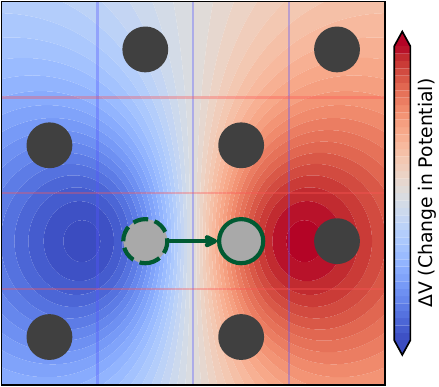}
    \caption{
   2D cross-section of the cubic lattice with a contour map showing the electrostatic potential change when a particle jumps to an adjacent site. Warm and cool colors indicate potential increases and decreases, respectively. }
    \label{fig:schematic_long_range}
\end{figure}

The critical challenge in applying kMC to systems with Coulomb interactions is calculating and updating the propensities. Because the electrostatic interaction is long-ranged, a single jump of a charged particle alters the energy landscape for all potential processes, requiring a global update of all propensities after every event. While each kMC event changes the system state only locally, i.e., in $\mathcal{O}(1)$, the number of affected propensities scales with the system size, leading to an at least $\mathcal{O}(N)$ update. This is depicted in Fig. \ref{fig:schematic_long_range}, which shows the change in electrostatic potential when the light grey charged particle moves from its initial position (dashed circle) to a new position (solid circle).

For the formulation of the propensities, we employ a Transition-State-Theory-type expression and the Brønsted-Evans-Polanyi (BEP) approximation \cite{BEP1938, Xiao2025AtomicScale} to incorporate interactions, which is closely related to the Miller--Abrahams description for particle hopping \cite{Coropceanu,Miller1960ImpurityCA}.
For the hopping process $\indx$, the propensity is the product of two factors: $\theta_\indx$, which is an indicator function representing local occupancy, and $f_\indx$, which is the rate of the process:
\begin{align}
 a_\indx(x)=& \ \theta_\indx(x)f_\indx(x),\label{eq:PropensityFunction} \\
 f_\indx(x)=& \ k_{\indx} \ \exp \left(-\beta \left(E_{0,\indx} + \alpha \Delta E_\indx(x)\right)\right).\label{eq:RateExpression}
\end{align}
$\theta_\indx(x)$ ensures that only feasible processes have non-zero propensities, i.e., it equals 1 if the process is possible and 0 otherwise. $k_\indx$ is the attempt frequency of the process $\indx$, $\beta$ is the inverse temperature, and $E_{0,\indx}$ is the barrier  of the process $\indx$ in the low occupation limit. $\Delta E_\indx(x)$ is the energy difference between the states caused by the process $\indx$. The parameter $\alpha$ determines how the transition state's energy depends on the initial and final states. Typically, $\alpha=1/2$ is considered for symmetric diffusion processes\cite{Gao_2022, Xiao2025AtomicScale}.
Note, Eq. \eqref{eq:RateExpression} is only senseful when the argument in the exponential function evaluates to less than zero. If this cannot be guaranteed, a correction can be applied to avoid negative barriers while maintaining microscopic reversibility \cite{Franziska_Hess}.
The energy difference $\Delta E_\indx(x)$ contains the effect of a constant applied electric field, $\boldsymbol{F}$ as well as of the Coulomb interactions. The electric field modifies the energy change associated with a particle jump in a specific direction. Therefore, the energy difference resulting from applying each process is calculated as
\begin{equation}\label{eq:Delta_energy}
\Delta E_\indx(x)=E(x+d_\indx) - E(x) - \sum\limits_i q_i \boldsymbol{F} \cdot \boldsymbol{\Delta R}_{\indx,i},
\end{equation}
where $q_i$ is the charge of the type of particle corresponding to the $i$-th entry in our state vector $x$ and $\boldsymbol{\Delta R}_{\indx,i}$ is the displacement of this charge in the process $\indx$. The Coulomb interaction energy $E(x)$ of the configuration $x$ can be expressed as $E(x)= x^TVx$, where $V_{ij}$ is the interaction potential between two particles at sites corresponding to the entries $i$ and $j$ in $x$. 

To address the computational bottleneck of updating propensities, we implement an exact reformulation that handles the recalculation needed after each process. This method involves no additional approximations but is achieved purely by reformulation of the propensity relation Eq. \eqref{eq:PropensityFunction} and Eq. \eqref{eq:RateExpression}.  After an executed process $\chi_n$ in step $n$, the energy difference for all processes $\indx$ can be updated as:
\begin{equation}\label{eq:EnergyUpdate}
\Delta E_\indx(x_{n+1})=\Delta E_\indx(x_{n}) + 2d_\indx^TV d_{\chi_n}.
\end{equation}
This allows us to update the function $f$ recursively:
\begin{align}
f_{\indx,n+1} &= f_{\indx,n} \cdot g_{\indx,\chi_n}, \label{eq:FactorUpdate} \\
\text{with} \quad
g_{\indx,\chi_n} &= \exp\left(-\beta \alpha \cdot 2 d_\indx^T V d_{\chi_n} \right),
\end{align}
where $f_{\indx,n+1}$ is the value of rate $f$ for the process $\indx$ in step $n+1$.
The matrix $g_{\indx,\chi}$ can be precomputed efficiently depending on the structure of the lattice and the interaction matrix and stored before simulation, dramatically reducing computational cost and making simulations of large systems feasible. Updating the propensity values is then achieved through two vector-vector multiplications: one for updating the factors $f$ and one for updating the propensities $a$, resulting in $\mathcal{O}(N_\text{proc.})$ computational complexity per step. 
For a system with $N$ lattice sites with local processes, there are $\mathcal{O}(N)$ number of possible processes at any given state leading to a linear complexity in system size per step.

While achieving $\mathcal{O}(N)$ computational complexity per step, a naive implementation would require $\mathcal{O}(N^2)$ storage for the matrix $g_{\indx,\chi}$. Even for relatively small problems, this might lead to intractable RAM requirements. However, for periodic lattices, processes can be grouped by their type, e.g., diffusion to the left, and their process vectors $d_\indx$ are identical up to translations on the lattice, i.e., we can assign an individual process to a unit cell of the lattice. The matrix $g_{\indx,\chi}$ will then only depend on the types of the processes $\indx$ and $\chi$, and the difference between the integer vectors identifying the respective unit cells. This convolutional structure allows for $\mathcal{O}(N)$ storage at the cost that the update rule Eq.~\eqref{eq:FactorUpdate} requires a little more bookkeeping. For non-periodic problems, e.g., amorphous materials, alternative approaches, such as hierarchical  matrix techniques\cite{Hackbusch_HierarchicalMA}, must be adapted for efficient storage of the matrix $g_{\indx,\chi}$.

\section{Results: cubic host lattice}
\begin{figure}
  \begin{minipage}{0.5\columnwidth}
    \includegraphics[width=\textwidth]{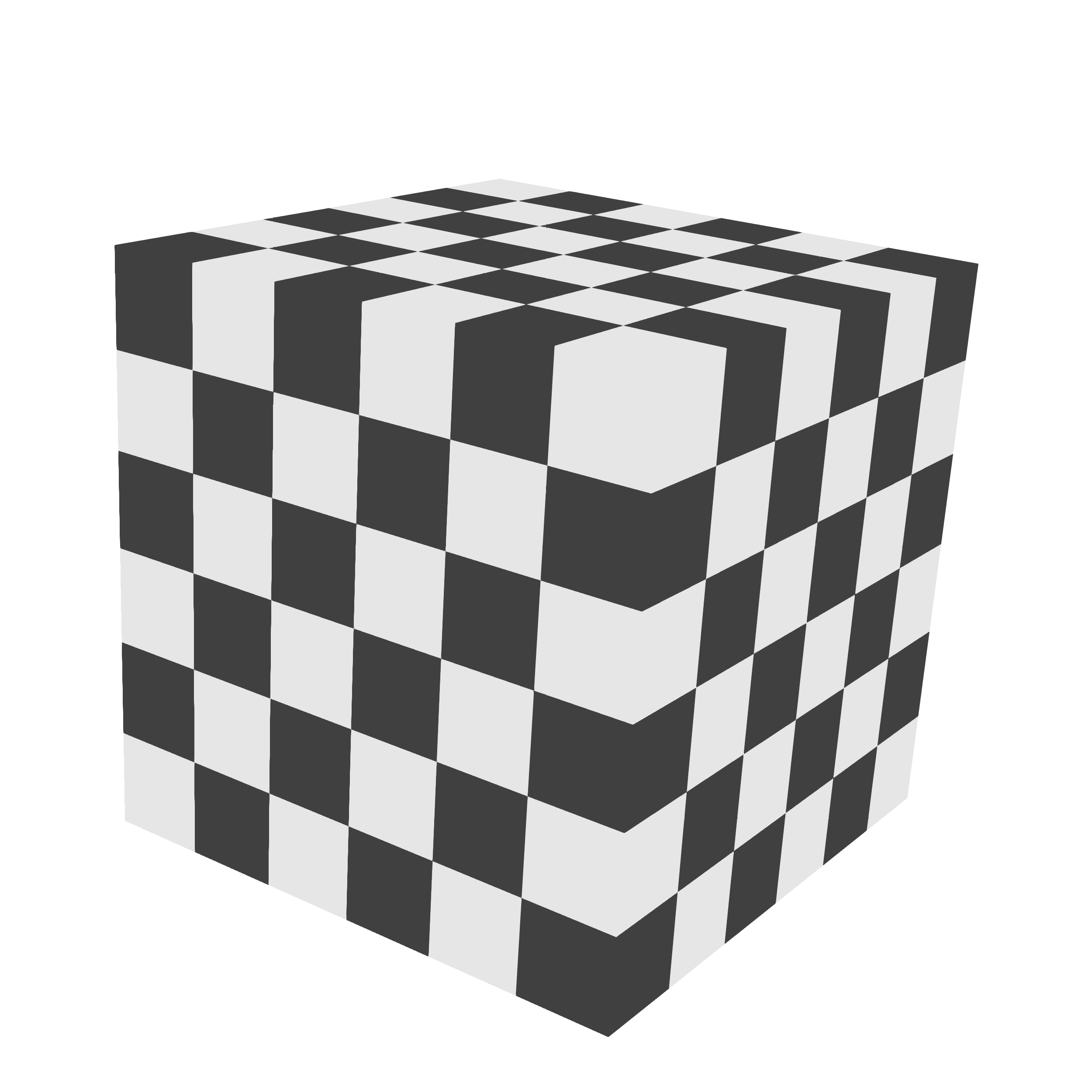}
  \end{minipage}
  \hfill
  \begin{minipage}{0.45\columnwidth}
    \captionsetup{width=\textwidth, justification=justified}
    \caption{
   Cubic lattice at $100$\,\% SOC showing intercalated ions arranged in a checkerboard pattern. Dark and light grey denote occupied and empty sites, respectively.}
    \label{fig:cubic_structure}
  \end{minipage}
\end{figure}

We illustrate the approach for an isotropic 3D cubic host lattice with a lattice spacing $\ell$.
We consider a single type of diffusing species with the charge $q$ per particle, and each unit cell has only a single site. Since each lattice site can carry only one particle, we choose the representation $x$ for the state such that $x_i=1$ corresponds to an occupied site and $x_i=0$ to an empty site. We further restrict to the simplest kind of elementary events, i.e., we restrict to particle hops to empty nearest-neighbor sites.
In our simulations, we apply an external electric field $F\boldsymbol{e}$, where $\boldsymbol{e}$ is a unit vector pointing in one lattice direction. We further restrict to cubic simulation boxes with $S$ unit cells in each direction, thus $N=S^3$ and choose $S$ to be a multiple of $2$.

Due to the isotropic cubic lattice, all hopping directions are equivalent, and all processes have the same (zero occupancy) activation energy $E_0$ and attempt frequencies $k_0$. This leads to a particularly simple parametric dependence for the  rate factors $f_\nu$
\begin{equation}\label{eq:ParametricFactorDependence}
 f_\nu(x)= k_p \exp(- \tfrac{1}{2} \ \lambda_T^{-1} (\tilde{E}(x+d_\nu)- \tilde{E}(x) - \delta_\nu\lambda_F)),
\end{equation}
where $\delta_\nu=1$ for hops in the field direction, $\delta_\nu = -1$ for hops in the opposite direction, and $\delta_\nu =0$ else. The function $\tilde{E}(x)$ is the interaction energy for a reference $S^3$ cubic lattice with unit dielectric constant, lattice spacing, and charge. It, thus, depends only on the size of the simulation box $S$ but not on any physical specifics of the problem at hand.
The common prefactor $k_p=k_0\exp(-\beta E_0)$ can be used to define the natural time unit $\tau:=k_p^{-1}$. Thus, the only nontrivial parameter dependence of the model is on $S$, SOC and the two dimensionless parameters
\begin{equation}\label{eq:lambda_T}
\lambda_T :=
\frac{\epsilon \ell k_\text{B}}{q^2} T,
\end{equation}
and
\begin{equation}\label{eq:lambda_F}
\lambda_F :=
 \frac{\epsilon \ell^2}{q}  F,
\end{equation}
where $k_\text{B}$ is the Boltzmann constant, $T$ is the temperature, $\epsilon$ is the dielectric constant, and  $q$ is the particle's charge. 
As revealed by Eq. \eqref{eq:ParametricFactorDependence} to \eqref{eq:lambda_F}, these two parameters, $\lambda_T$ and $\lambda_F$ have the meaning of temperature and field strength and can be adjusted independently by both.

We investigate the stationary non-equilibrium response under an applied electric field as a function of state of charge, temperature, and field strength. In particular, we will investigate the high-SOC regime, where strong Coulomb interactions and configurational constraints are expected to significantly influence transport. We define situations with $N/2$ charge carriers in the simulation box as $100$\,\% SOC ($50$\,\% physical occupation). This is motivated by the symmetry of the problem where the case with $M<N$ particles is equivalent to the case with $N-M$ particles. We consider $\lambda_T \in [0.007,0.010]$ and $\lambda_F\in [0.054,0.108]$.
These ranges correspond to variations of the temperature $T$ within $[300,440]$\,K and fields between $[0.082,0.162]$\,V/\AA {} for hosts with $q=-e$, $\ell=2.46$\,\AA, and a relative dielectric constant of $20$, which roughly agrees with the values in highly lithiated graphite as an important battery material\cite{ANNIES2023141966}.
In the following, we report parameter settings in terms of the dimensionless parameters $\lambda_T$ and $\lambda_F$ alongside the values for $T$ and $F$ for these material parameters, for clarity. 
The applied fields are relatively large, and it would be, at least, very demanding to experimentally realize such fields on macroscopic scales. However, such high electric fields can easily appear on nanometer scales, e.g., at dielectric interfaces\cite{double_layer_Elliott, Montenegro2021}.
Linear response might not apply in such situations, and working with different high field values provides insight into transport mechanisms in the nonlinear-response regime.
Simulations have primarily been performed on cubic boxes of size $S=12$, but systematic size-variation studies have been conducted to exclude finite size effects (see Fig. \ref{fig:size_effect}).
The initial configuration for $100$\,\% SOC ($50$\,\% physical occupancy) is a three-dimensional checkerboard arrangement as displayed in Fig. \ref{fig:cubic_structure} where black and light grey cells correspond to empty and occupied sites, respectively. This configuration minimizes the electrostatic energy and thus reduces the relaxation time to reach steady state. Initial configurations for lower SOCs are created in a sequential way, where a particle is removed from the initial configuration with one particle more such that the electrostatic energy is minimal. For larger SOCs, initial conditions are equivalently constructed sequentially, replacing vacancies with charge carriers. Steady-state observables are obtained by time-averaging after a long-enough relaxation period to ensure the system has reached steady state. For both, we employ $10^8$ kMC steps.

\subsection{Computational Performance}

\begin{figure}[h]
\includegraphics[width=0.9\columnwidth]{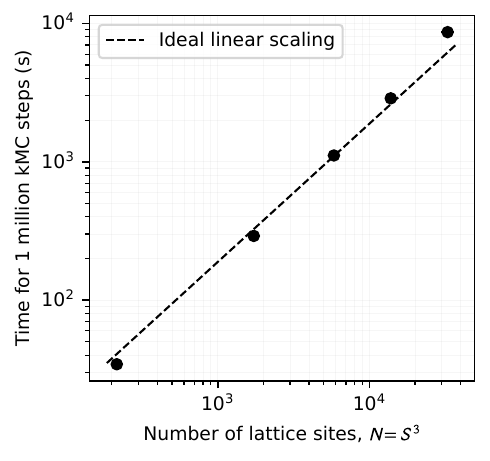}
\caption{Computational performance of the kMC implementation as a function of lattice size.
The symbols show the average wall-clock time required to perform 1 million kMC steps on cubic lattices containing $N = S^3$ sites, with $S = [6, 12, 18, 24, 32]$.
The dashed line indicates the ideal ($\mathcal{O}(N)$) complexity.
}
    \label{fig:computational_performance}
\end{figure}

To assess the computational efficiency of the method, we measured the wall-clock time required to perform 1 million kMC steps for cubic lattices of increasing size using our in-house kMC package \cite{BSSST}. The benchmarks were performed on a single Intel Xeon IceLake-SP 8360Y CPU core of the Raven supercomputer at the Max Planck Computing and Data Facility (MPCDF).
Figure~\ref{fig:computational_performance} shows the resulting computational time as a function of the total number of lattice sites, $N=S^3$ (black dots) for an SOC of $100$\,\%, $\lambda_T=0.009$ ($T=365$\,K) and $\lambda_F=0.078$ ($F=0.118$\,V/AA).
The measured runtime increases from approximately 34 seconds for $S=6$ to approximately 144 minutes for $S=32$ closely following the ideal scaling behavior $\mathcal{O}(N)$ which is indicated by the dashed line. The slight deviation from ideal linear scaling becomes more apparent for the largest lattice considered ($S=32$). This behavior is likely caused by hardware-level limitations such as increased memory-access overhead and cache or memory-bandwidth saturation at larger system sizes, rather than an intrinsic algorithmic bottleneck. Nevertheless, the observed scaling remains close to $\mathcal{O}(N)$, indicating that the method remains computationally tractable even for large systems with fully resolved Coulomb interactions.

\subsection{Parametric Dependence of the Current Density}

\begin{figure*}[t] 
\centering 
\begin{subfigure}[t]{0.45\textwidth} 
\centering 
\includegraphics[width=\linewidth]{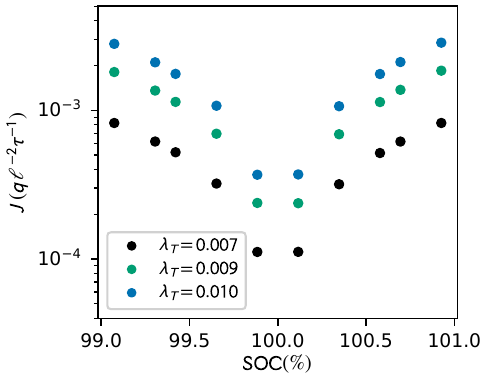} 
\caption{} 
\label{fig:current_density_T} 
\end{subfigure} 
\begin{subfigure}[t]{0.45\textwidth} 
\centering
\includegraphics[width=\linewidth]{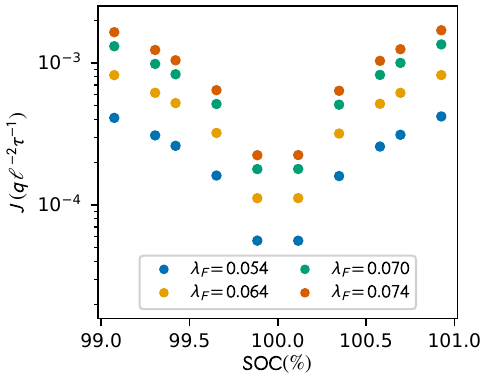} 
\caption{} 
\label{fig:current_density_F} 
\end{subfigure}
\caption{Current density, $J$, as a function of SOC. 
(a) Temperature dependence at a fixed electric field of $\lambda_F = 0.064$ ($F=0.096$\,V/\AA) for three temperatures $\lambda_T = \{0.007,0.009,0.010\}$ ($T=\{300,365,440\}$\,K). The minimum values at $\mathrm{SOC}=100$\,\% fall below the displayed plotting range and are therefore omitted for clarity.
(b) Dependence on the applied electric field strength at fixed temperature $\lambda_T=0.007$ ($T=300$\,K). Different colors correspond to different applied electric fields.}
\label{fig:current_density}
\end{figure*}

The central objective in charge transport is the current density and its response to changes of the system's parameters, which are the SOC, temperature $T$ ($\lambda_T$) and field $F$ ($\lambda_F$) for the considered problem class. To investigate this dependence, we first applied  a fixed electric field at different temperatures and simulated the stationary current for SOCs in the range $[99,101]$\,\% .  Figure~\ref{fig:current_density_T} displays the results in dependence on the SOC for $\lambda_F=0.064$, i.e., an electric field of $F=0.096$\,V/\AA. As can be expected,  higher temperature implies higher current for a given SOC, and comparatively small temperature variations lead to significant current variations. For all temperatures, we further observe a strong dependency on small variations of the SOC with a pronounced minimum at $\mathrm{SOC}=100$\,\%. The minimum values at this point are $J=3\times10^{-9}$\,$q\ell^{-2} \tau^{-1}$ for $\lambda_T=0.007$ ($T=300$\,K), $J=7\times10^{-8}$\,$q\ell^{-2} \tau^{-1}$ for $\lambda_T=0.009$ ($T=365$\,K), and $J=10^{-5}$\,$q\ell^{-2} \tau^{-1}$ for $\lambda_T=0.010$ ($T=440$\,K).
As the SOC deviates from $100$\,\%, either through the addition or removal of charge carriers, $J$ increases rapidly and symmetrically with respect to $100$\,\% SOC. This symmetry reflects the equivalence of charged particles and vacancies in the lattice.
Figure~\ref{fig:current_density_F} shows the same dependence of $J$ but now for four different applied fields with the same temperature of $\lambda_T=0.007$ ($T=300$\,K).
For fixed SOC, we observe the expected behavior with stronger fields leading to higher currents. These results also show that we are beyond the linear-response regime as the applied fields differ by about $35$\,\% but the resulting currents by a factor $3.9$. 
At all field strengths, the pronounced minimum at $100$\,\% SOC is present, and the currents are about $6$ orders of magnitude smaller than at $99$\,\% SOC.

\begin{figure}[h]
\includegraphics[width=0.9\columnwidth]{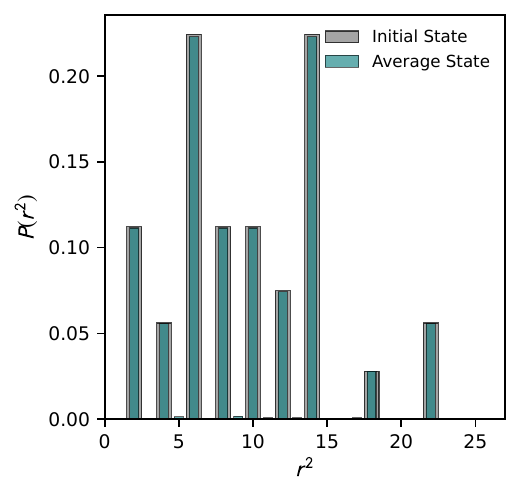}
\caption{\label{fig:distance_distribution_locking}
Normalized pairwise squared-distance distribution between ions for the initial checkerboard configuration and the time-averaged state exposed to an electric field $\lambda_F=0.078$ ($F=0.118$\,V/\AA) at $100$\,\% SOC and a temperature $\lambda_T=0.009$ ($T=365$\,K).
}
\end{figure}

\section{Discussion}

The results in Figs. \ref{fig:current_density_T} and \ref{fig:current_density_F} show a pronounced minimum in $J$ at $100$\,\% SOC, reflecting strong immobilization and locking of the system. Besides the symmetry, these results provide an additional motivation for the labeling of $50$\,\% occupancy as $100$\,\% SOC. As we charge the system, the process will slow down by orders of magnitude when we approach $50$\,\% occupancy due to the very small current there. From the outside, this will look as if we have reached the maximum SOC, simply because it would take an extremely long exposure to elevated fields to cross this point. The strong variation of the current close to  $100$\,\% SOC might also serve as one explanation for why mobility measurements in battery materials often show a wide spread of reported values in this regime\cite{Cristina2023,Levi_1997}.
Little variations of the charge carrier density around $100$\,\% SOC can lead to largely different currents.

\textbf{Locking at $\mathbf{100}$\,\% SOC:} 
To first understand the origin behind the locking, we analyze the structural evolution of the system  at $100$\,\% SOC. Figure~\ref{fig:distance_distribution_locking} compares the normalized pairwise squared-distance distributions between the initial checkerboard configuration and the time-averaged state under the applied field for $\lambda_T=0.009$ ($T=365$\,K) and $\lambda_F =0.078$ ($F=0.118$\,V/\AA). Both distributions are nearly identical, indicating that the system remains close to its initial configuration.  A quantitative analysis of the simulation trajectories further shows that the system remains in the checkerboard structure for more than $96$\,\% of the simulation time. For lower temperatures, the checkerboard structure becomes more dominant with more than $99$\,\% of the time spent in this configuration for $\lambda_T=0.007$ ($T=300$\,K). For higher temperatures, the checkerboard structure starts ``dissolving'' with $80$\,\% of the time spent in checkerboard configuration for $\lambda_T=0.010$ ($T=440$\,K) and
$\lambda_F =0.080$ ($F=0.120$\,V/\AA) and only $60$\,\% in the same temperature and $\lambda_F =0.094$ ($F=0.142$\,V/\AA).

\begin{figure}[h]
\includegraphics[width=\columnwidth]{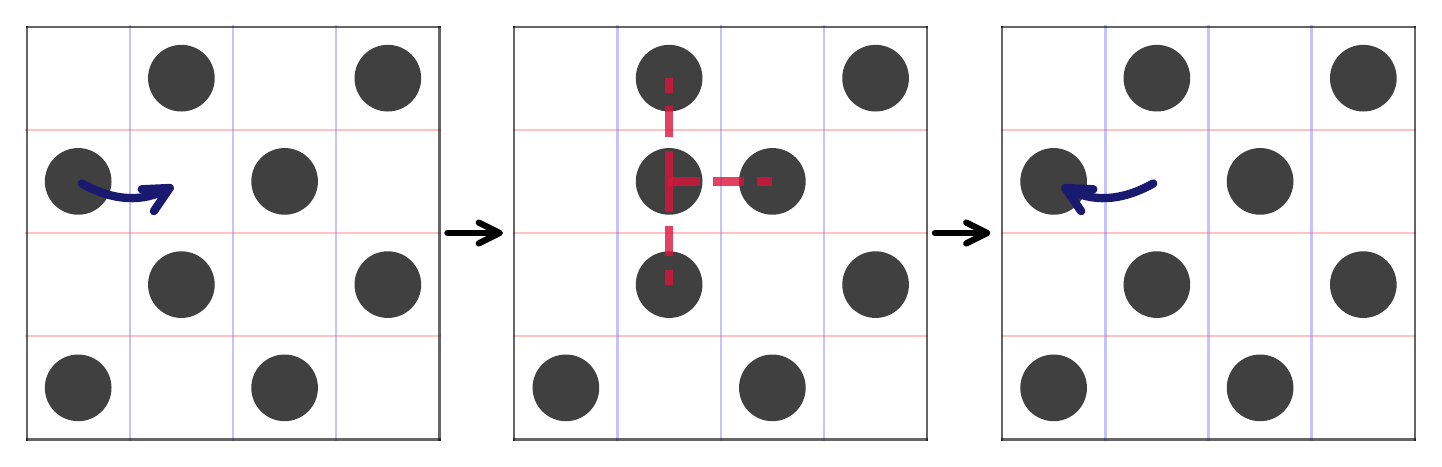}
\caption{\label{fig:locking_schematic}
The checkerboard structure in ionic lattices is a highly stable structure at low temperatures. The left panel shows the initial checkerboard configuration with minimal electrostatic energy. The middle panel  depicts the particle jump to a nearest neighbor site. This jump creates an unfavorable configuration with strong Coulomb repulsion (dashed lines) from other ions. This repulsion drives the particle back to its original position, maintaining the stable checkerboard pattern.}
\end{figure}

This immobilization, even in the presence of an externally applied electric field, is due to the absence of possible jumps with low energy penalty. Particles attempting to move to the neighboring sites face large repulsions from other particles and therefore tend to jump back to their previous position, as illustrated in Fig. \ref{fig:locking_schematic}. However, removing even a small number of carriers breaks this symmetry, creating low-energy pathways that enable transport. This explains the abrupt increase in current density, $J$, observed upon deviation from $100$\,\% SOC shown in  Fig. ~\ref{fig:current_density}.

\begin{figure}[h]
\includegraphics[width=\columnwidth]{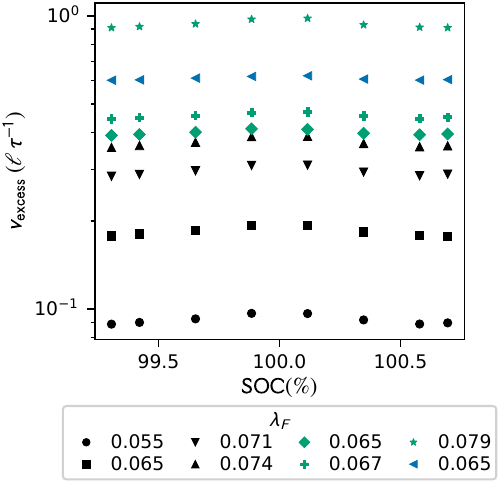}
\caption{\label{fig:excess_velocity}
Average excess carrier velocity, $v_{\mathrm{excess}}$, as a function of SOC for different temperatures (black: $0.4$, green: $0.5$, blue: $0.6 \ \Tunit$) and applied field strengths (different symbols). The excess velocity remains approximately constant, demonstrating linear scaling of conductivity with excess carrier population.
}
\end{figure}

\textbf{Defect transport:}
The shown results and interpretation indicate a fundamental change in the transport mechanism by introducing defects—either excess charges or vacancies— into this Coulomb superlattice at 100\% SOC. At low defect concentrations, we would expect to have two kinds of transport mechanisms: i) a global transport in the superlattice, which is also there at 100\% SOC and is largely unaffected by the small number of defects; and ii) a defect-driven transport mechanism operating on the superlattice. Accordingly, we decompose the total current density into two contributions: a background term, $J_{100\%}$, corresponding to the current density of the superlattice at $100\%$ SOC  (depending only on field and temperature), and an excess current density, $J_{\mathrm{excess}}(\mathrm{SOC})$, arising from defect transport and dependent on the state of charge. In other words, we define $J_{\mathrm{excess}}:= J - J_{100\%}$.
To clarify the nature of the defect-driven  mechanism, we introduce the average excess charge carrier velocity, $v_{\mathrm{excess}}$, defined as $J_{\mathrm{excess}}$ divided by the defect density which would be proportional to $\Delta \mathrm{SOC}:=\mathrm{SOC}-\mathrm{SOC}_{100\%}$. Thus, while the total SOC is the relevant normalization for the transport of $100$\,\% SOC (and at very low SOC), the appropriate normalization for $J_{\mathrm{excess}}$ is $\Delta SOC$ in the vicinity of $100$\,\% SOC. 

\begin{figure}[h]
\includegraphics[width=\columnwidth]{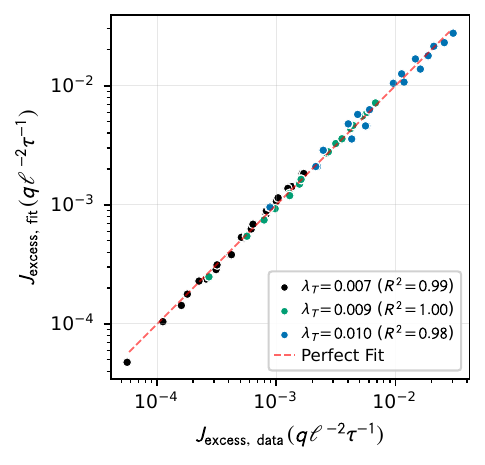}
\caption{\label{fig:parity_ultra_dilute}
Parity plot validating the ultra-dilute transport model (Eq.~\eqref{eq:J}). Three parameters are fitted using data from different temperatures, field strengths, and SOCs.
}
\end{figure}

As shown in Fig. \ref{fig:excess_velocity}, the excess velocity remains approximately constant across different SOC values. This means conductivity scales linearly with excess carrier population, with each carrier contributing equally to transport regardless of overall carrier density. 
This equal contribution of the defects in the transport is characteristic  for the ultra-dilute, non-interacting defect regime. Assuming an independent lattice model for the defects, the  transport then follows mean-field behavior, where contributions from excess carriers are additive. The excess current density can then be expressed as 
\begin{equation}
    \label{eq:J_general}
    J_{\mathrm{excess}} = c\ (f_{+}-f_{-})\ \Delta \mathrm{SOC},
\end{equation}
where $f_{\pm}$ are the effective rates of defect movements in the direction of the applied field ($f_{+}$) and opposite to it ($f_{-}$).
To incorporate field and temperature dependence, we model the rates in the same way as for the full model Eq.~\eqref{eq:RateExpression} but without interaction,
i.e. $f_{\pm} \propto \exp(-\lambda_T ^{-1}(E_{\text{def.}} \mp \tfrac{1}{2} Q\lambda_F)$. Here, $E_{\text{def.}}$ is the effective barrier and $Q$ is a charge factor encoding the difference between the effective defect charge and the charge $q$ of the particles in the full model. For the current, we arrive at the expression
\begin{equation}
    \label{eq:J}
    J_{\mathrm{excess}} = A\Delta \mathrm{SOC} \exp(-\lambda_T^{-1} {E}_{\text{def.}}) \sinh(\tfrac{1}{2} \lambda_T^{-1} \lambda_F Q ).
\end{equation}
We fit the parameters $A$, ${E}_{\mathrm{def.}}$ and $Q$ in Eq.~\eqref{eq:J} using our simulation data across different temperatures, applied fields, and states of charge. Since $J_{\mathrm{excess}}$ spans several orders of magnitude, the fit is performed on the logarithm of Eq.~\eqref{eq:J}. The fit yields $R^2 = 0.996$, with $A = 0.013 \pm 0.001 q\ell^{-2} \tau^{-1}$, an activation energy $\tilde{E}_{\text{def.}} = 0.0655 \pm 0.0005 $ (approximately \(0.24\,\mathrm{eV}\) for the physical parameter mapping considered here), and a charge factor $Q = 1.132 \pm 0.017$. Figure~\ref{fig:parity_ultra_dilute} shows parity plots comparing simulated and predicted excess current densities for all three temperatures, demonstrating excellent agreement.

At low temperatures, the background current density, $J_{100\%}$ is negligible, but at $\lambda_T = 0.010$ ($T=440$\,K), $J_{100\%}$ is significant, and therefore using $J$ in the model instead of $J_{\mathrm{excess}}$  leads to noticeable deviations. These deviations indicate the onset of collective transport beyond the dilute-defect regime driven by thermal disruption of the checkerboard ordering. This is to be expected from the analysis of the configuration at $100$\,\% SOC where, at $\lambda_T=0.010$ ($T=440$), the checkerboard structure is maintained for only $60$\,\% of the simulated time, compared to more than $99.95$\,\% at $\lambda_T=0.007$ ($T=300$).

\begin{figure}[h]
\includegraphics[width=\columnwidth]{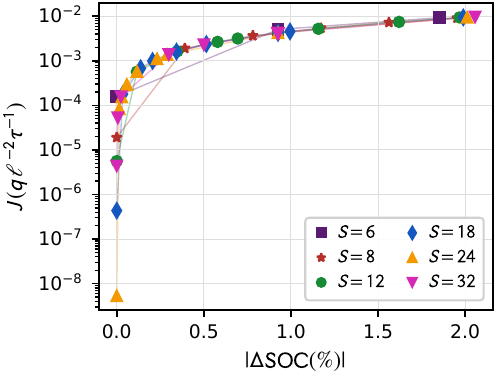}
\caption{\label{fig:size_effect}
Current density as a function of state of charge deviation at $\lambda_T=0.009$ ($T=365$\,K) and an applied electric field $\lambda_F=0.078$ ($F=0.118$\,V/\AA) for five system sizes (S = 6, 8, 12, 18, 24, 32), corresponding to cubic lattices of side length S. Each symbol is for a specific system size, and the lines are connecting symbols only for eye guidance. At $100$\,\% SOC, the current density in different sizes does not match and span over a wide range, but as long as defects are introduced, the results in all sizes follow the same trend.
}
\end{figure}

\textbf{Finite size effects:} The extreme drop of current density at $100$\,\% SOC in low temperatures and the sharp increase in its surroundings reflect the checkerboard ordering and transition between two transport regimes: excess-carrier transport for $\Delta \mathrm{SOC}>0$ and vacancy-mediated transport for $\Delta \mathrm{SOC}<0$. By our analysis, we also expect that the derivative of the current with respect to the SOC will be discontinuous at $100$\,\% SOC. This implies that the $100$\,\% SOC points will be - at least near - a second order phase transition and, thus, finite size effects must be critically tested for. We therefore have conducted systematic tests varying $S$ in the range $[6,32]$. Figure~\ref{fig:size_effect} exemplarily shows the results for different system sizes for the case $\lambda_T=0.009$ ($T=365$\,K), $\lambda_F=0.078$ ($F=0.118$\,V/\AA) and varying SOC. Our results contain only minor quantitative finite size effects except at exactly $100$\,\% SOC, where we cannot get converged results. This is to be expected, as $100$\,\% SOC would be exactly at the phase transition where correlation length and time will diverge.

\section{Conclusion}
We have introduced a kinetic Monte Carlo update procedure for systems with long-range interactions and demonstrated its use for charge carrier diffusion on a simple cubic host lattice. The method allows to include full-range Coulomb interactions without recomputing all process rates from scratch after each event. The resulting simulations show the predicted linear scaling with system size, making large interacting kMC simulations computationally tractable.

Applied to the cubic host-lattice model at finite fields, the method reveals that the system forms an ordered superlattice at $100$\,\% SOC for the considered temperatures, which leads to a strong suppression of charge transport. Small deviations from this ordered state, either by adding particles or creating vacancies in the superlattice, introduce mobile defects and produce a rapid increase in the current density. Close to $100$\,\% SOC, the transport of these defects can be well described by a simple non-interacting particle model in the ultra-dilute regime.

Although the numerical results presented here are specific to a simple cubic host lattice with one diffusing species, the mechanisms identified are not tied to this particular geometry, and we expect analogous locking and dilute-defect transport regimes to occur in more complex host structures. Future work will therefore utilize the proposed methodology for simulating charge transport in more complex host structures, such as graphite \cite{Chiara2021, ANNIES2023141966}, but, also for addressing the effects of imperfections in the host lattice.

\begin{acknowledgments}
This work was partially supported by Germany’s Excellence Strategy MATH+: Berlin Mathematics Research Center (EXC 2046). 
The authors want to thank Christian Carbogno and Sandra D\"opking, as well as the HPC Application Support team of the Max Planck Computing and Data Facility for their help with the implementation of the software.
\end{acknowledgments}
\bibliography{references_Sebastian}

\end{document}